\documentstyle[12pt]{article}
\newcommand{\be}{\begin{equation}}
\newcommand{\ee}{\end{equation}}
\newcommand{\bea}{\begin{eqnarray}}
\newcommand{\eea}{\end{eqnarray}}

\begin{document}
\title{Shape anisotropy and magnetic domain structures in striped monolayers}
\author{Paolo Politi$^1$ and Maria Gloria Pini$^2$}
\maketitle
\begin{center}
$^1$Dipartimento di Fisica dell'Universit\`a degli Studi di Firenze and
Sezione INFM, L.go E. Fermi 2, I-50125 Firenze, Italy\\
$^2$Istituto di Elettronica Quantistica, Consiglio Nazionale delle
Ricerche, Via Panciatichi 56/30, I-50127 Firenze, Italy
\end{center}
\begin{abstract}
We study the effect of dipolar interactions on a magnetic striped monolayer
with a microscopic unit cell of square symmetry, and 
of size $(N_x\times N_y)$ spins.
Even if the aspect ratio $r=N_x/N_y$ is very large,
an in-plane shape anisotropy is always negligible, except if $N_y$ 
is fairly small $(N_y<40)$. In-plane domains are not possible, 
except for values of the dipolar coupling larger than the domain 
wall energy.
\end{abstract}

\section{Introduction}

An important and well known feature of magnetic dipole-dipole interactions
is that, in spite of their weakness with respect to exchange coupling, 
they play an important role in magnetic systems~\cite{Aharoni}.  
In fact, the long range character of the magnetostatic interactions 
is relevant for determining both the ground state of the system~\cite{Kittel} 
and the excitation spectrum~\cite{HP}; 
perhaps, the most striking consequence of dipolar interaction is
the breaking of a bulk sample in several magnetic domains.

The peculiarity of dipolar interactions in three dimensions 
is elucidated by the fact that a shape anisotropy is always present, 
independently of the size of the sample: 
as a consequence of this, it has no meaning to speak about an ``infinite
sample" without specifing its limiting shape.  
For example, for the slab geometry (see Fig. 1b), 
the demagnetizing factors are 0 in the $x,y$ directions
and 1 in the $z$ one, perpendicular to the slab. 
Thus, the in-plane collinear 
state has a lower energy than the perpendicular collinear one, whose 
surface energy density turns out to be proportional to the 
thickness of the slab, and therefore the shape anisotropy 
{\it per spin} of a slab is a constant and 
doesn't depend on its thickness $N_z$!

In genuine two dimensional magnetic systems, like magnetic
films of atomic thickness (Fe, Co, Ni, Gd) grown on nonmagnetic 
substrates (Au, Ag, Cu, $\dots$), the situation is quite different: 
the shape anisotropy due to magnetic dipole-dipole interactions 
favours an in-plane magnetization without establishing any preferential 
direction within the film~\cite{Maleev,YKG},
and the magnetization direction of a collinear configuration
is determined by the competition 
with possible easy-axis anisotropies generated by the breaking of the 
translational invariance in the growth direction and which may favour 
a perpendicular state. For in-plane magnetization, a monodomain 
configuration is energetically favoured; for perpendicular
magnetization, the system prefers to break into
domains~\cite{Allenspach,Politi}, however weak the dipolar interaction may be.

Recently, two-dimensional mesoscopic structures~\cite{art_exp}
have attracted attention for their potential technological applications.
They are characterized (see Fig.~1a) by a ``small" lateral dimension
in one direction ($N_y\ll N_x$: magnetic wires) or in both
directions (magnetic dots). Such structures will be called
{\it striped monolayers} in the following, and $N_x,N_y$ will be
allowed to assume any value.

The aim of the present paper is to
elucidate the role played by magnetic dipole-dipole interactions in
such systems, provided that the magnetization lies in the plane of 
the stripe and the microscopic unit cell has square symmetry,
and to ascertain whether the results valid for a slab 
apply analogously, namely: (i) the presence of a strong shape
anisotropy favouring the $x$ direction if $N_x \gg N_y$, and
(ii) the existence of domains, if some in-plane magnetocrystalline
anisotropy forces the magnetization along the $y$ direction.
As we will show, neither of the previous points holds. In
contrast, we find that
(i)~even for a stripe with an infinite aspect ratio
$(N_x/N_y \to \infty)$, the shape anisotropy {\it per spin} vanishes 
upon increasing $N_y$, and 
(ii)~in-plane domain structures can appear
only for very large values of the dipolar interaction $\Omega$.

\section{Shape anisotropy of a striped monolayer}
Let us start by writing down the dipolar energy for an in-plane
collinear ground state:
\be
E_{dip} = {\Omega\over 2}\sum_{\vec n,\vec m} {1\over r_{nm}^3}
\left[ 1-3{(\vec S\cdot\vec r_{nm})^2\over r_{nm}^2}\right]~,
\label{E_dip}
\ee
where the distance $r_{nm}=|\vec n-\vec m|$ (with 
$\vec n \ne \vec m$) is measured in units of the 
square lattice constant.

The shape anisotropy is determined
solely by the anisotropic part of $E_{dip}$ (the second term
in square brackets), which becomes, if $\vec S=(\cos\theta,\sin\theta)$
and up to a constant:
\be
E_{dip} = -3\Omega\cos^2\theta\cdot{1\over 2}\sum_{\vec n,\vec m}
{(x_{nm}^2-y_{nm}^2)\over r_{nm}^5}\equiv -3\Omega\cos^2\theta\cdot{\cal S}~,
\ee
where the ``cross" term of the form $\sum_{\vec n,\vec m}(x_{nm}y_{nm}/
r^5_{nm})$ vanishes if the system has at least one symmetry axis (a
rectangular monolayer has two).

The previous expression vanishes for an infinite two dimensional
monolayer, or for any square-shaped system ($N_x=N_y$), provided
that the symmetry of the microscopic unit cell is square.
For a rectangle with $N_x>N_y$, a couple of points $(\vec n,\vec m)$
will generally correspond to a value $|x_{nm}|>|y_{nm}|$, so that the
summation ${\cal S}$ will have a strictly positive value, 
$E_{dip}$ will be minimal for $\theta=0$ and the
magnetization will be oriented along
the $\hat x$ axis. 

The relevant quantity is the dipolar energy per spin: 
$\epsilon_{dip}=E_{dip}/N_xN_y$, and we are interested in analyzing the
dependence of $\epsilon_{dip}$ on $N_x$ and $N_y$. For $N_y=1$, in the limit
of an infinite chain of spins, it is immediately found that
\be
\epsilon_{dip}(N_x=\infty,N_y=1)=
-3\Omega\cos^2\theta\cdot{1\over 2}\sum_{n\ne 0} {1\over n^3}\equiv
-3\Omega\cos^2\theta\cdot\zeta(3)~,
\ee
where $\zeta(3)=1.2021$ is the Riemann's zeta function~\cite{Abramowitz}.

The competition between the dipolar contribution ($3\zeta(3)\Omega$) 
and possible in-plane 
ani\-so\-tro\-pies favouring the $\hat y$ axis, determines the actual 
direction of the magnetization.
In the following, we will evaluate $\epsilon_{dip}$ in a continuum 
approximation, for any value of $N_x,N_y$.

The exact definition of ${\cal S}$ is:
\be
{\cal S}\equiv {1\over 2}
\sum_{l,l'=1}^{N_x}\sum_{m,m'=1}^{N_y}
{ (l-l')^2 - (m-m')^2 \over
\left[ (l-l')^2 + (m-m')^2 \right]^{5/2}}
~~~\hbox{with}~~(l,m)\ne(l',m')~.
\label{S1}
\ee

It is useful to introduce the new integer variables: $x=(l-l')$, with
$-(N_x-1)\le x\le (N_x-1)$ and $y=(m-m')$, with
$-(N_y-1)\le y\le (N_y-1)$. ${\cal S}$ now writes as a sum on solely
two indices:
\be
{\cal S} = {1\over 2} \sum_{x=-(N_x-1)}^{N_x-1}
\sum_{y=-(N_y-1)}^{N_y-1} (N_x-|x|)(N_y-|y|)
{x^2 - y^2 \over (x^2 + y^2)^{5/2} }
\ee

At this point we exploit the fact that ${\cal S}\equiv 0$ if $N_x=N_y$.
Therefore, we will write $N_x=N_y + (N_x - N_y)$:
\bea
{\cal S}&=&{1\over 2}\sum_{|x|,|y|\le(N_y-1)} (N_y-|x|)(N_y-|y|)
{x^2 - y^2 \over (x^2 + y^2)^{5/2} }\nonumber\\
&+& {1\over 2}\sum_{|x|>(N_y-1)}^{|x|\le(N_x-1)}
\sum_{|y|\le(N_y-1)} (N_x-|x|)(N_y-|y|)
{x^2 - y^2 \over (x^2 + y^2)^{5/2} }\nonumber\\
&+& {1\over 2}(N_x-N_y)\sum_{|x|,|y|\le(N_y-1)} (N_y-|y|)
{x^2 - y^2 \over (x^2 + y^2)^{5/2} }\nonumber
\eea

The first sum vanishes, as seen by interchanging the two dumb indices
$(x,y)$; in the third sum, for the same reason we can get rid of the
quantity proportional to $(N_x-N_y)N_y(x^2-y^2)/(x^2 + y^2)^{5/2}$,
whilst in the other one the term $y=0$ does not contribute.
Finally, by using the parity of the addenda, we obtain:
\bea
{\cal S}&=& \sum_{x=N_y}^{N_x-1} \sum_{y=-(N_y-1)}^{N_y-1}
(N_x-x)(N_y-|y|) {x^2 - y^2 \over (x^2 + y^2)^{5/2}}\nonumber\\
&-& (N_x-N_y) \sum_{y=1}^{N_y-1} \sum_{x=-(N_y-1)}^{N_y-1}
y {x^2 - y^2 \over (x^2 + y^2)^{5/2} }
\label{S}
\eea

It is noteworthy that the previous expression is exact:
its use for a numerical calculation of ${\cal S}$ requires the
evaluation of $2N_xN_y$ terms, whilst Eq.~(\ref{S1}) demanded $N_x^2N_y^2$
terms!

The evaluation of ${\cal S}$ in the continuum approximation
$(\sum_{x,y}\rightarrow\int\int dxdy)$ is performed in the Appendix~A.
Here we will discuss the results.
First of all, let us consider the case of an infinite aspect ratio
($r=N_x/N_y=\infty$ with $N_y$ finite). 
The quantity ${\cal S}$ per spin is (see Eq.~(\ref{a1})):
\be
{{\cal S}\over N_x N_y} = {2\over 3N_y}\ln N_y + {c_1\over N_y}
\label{S2}
\ee

An important feature of the previous expression immediately comes out:
the shape anisotropy (per spin) vanishes upon increasing $N_y$, even if
$N_x/N_y=\infty$! This means that the shape anisotropy of an infinite
stripe ($N_x=\infty$) becomes rapidly negligible, as $N_y$ increases.

The numerical value of $c_1$ cannot be determined by our ``zero-order"
continuum approximation; in fact, the first-order correction,
given by the term in square brackets of
the Euler-Maclaurin summation formula (see Ref.~\cite{Abramowitz}):
\be
\sum_{x=a}^{b} f(x) = \int_a^b dx f(x) + {1\over 2}[f(a) + f(b)] + \dots
\ee
contributes just to the term of order $(1/N_y)$ in (\ref{S2}).
Since the leading term ($\approx (1/N_y)\ln N_y$)
dominates only logarithmically, the
constant cannot be neglected. In Eq.~(\ref{S2}), $c_1$ plays the role of
the anisotropy for a single line of spins:
\be
c_1 = \left. {{\cal S}\over N_x N_y}\right|_{N_y=1} = 
\sum_{l=1}^{\infty} {1\over l^3} = \zeta(3)~.
\ee

So, we will rewrite (\ref{S2}) in the form:
\be
{{\cal S}\over N_x N_y} = {2\over 3N_y}\ln N_y + {\zeta(3)\over N_y}~.
\label{S3}
\ee

In the limit $N_x=\infty$, the numerical calculation of ${\cal S}/N_x$
can be made much more efficient by means of the Ewald's summation technique~%
\cite{Ewald}, which allows to rewrite ${\cal S}/N_x$ as an exponentially
converging sum. This is done in Appendix~B. We are therefore able to
compare (see Fig.~2) the exact numerical result with the 
analytic approximation~(\ref{S3}). 
Even our ``zero-order" continuum approximation
gives a fairly good approximation. 

The following considerations are meant to corroborate further on
the previous results.
If $N_x=\infty$, we can exploit the translational invariance in the $\hat x$
direction to write down:
\bea
{{\cal S}\over N_x} &=& {1\over 2} \sum_{m,m'=1}^{N_y}\sum_{l=-\infty}
^{\infty} 
{ l^2 - (m-m')^2 \over \left[ l^2 + (m-m')^2 \right]^{5/2} }
\equiv {1\over 2}\sum_{m,m'=1}^{N_y} S(m-m')\nonumber\\
&=& {1\over 2}N_y S(0) + \sum_{c=1}^{N_y-1} (N_y-c)S(c)\nonumber
\eea

$S(c)$ represents the ``interaction" per unit length of two
lines at distance $c$, whilst $S(0)/2$ is the self-interaction of a line.
In the limit where $N_y$ can be treated as a continuous variable:
\bea
{\partial\over\partial N_y}\left({{\cal S}\over N_x}\right) &=& {S(0)\over 2} 
+\sum_{c=1}^{N_y} S(c)\nonumber\\
{\partial^2\over\partial N_y^2}\left({{\cal S}\over N_x}\right) 
&=& S(N_y)\nonumber
\eea

If the continuum approximation is applied also to the $\hat x$
direction, for $c\ne 0$ we will have:
\be
S(c) = \int_{-\infty}^{+\infty} dx {x^2 -c^2 \over (x^2 + c^2)^{5/2} }
= -{2\over 3c^2}
\ee
So, 
$$
{\partial^2\over\partial N_y^2}\left({{\cal S}\over N_x}\right)=
-{2\over 3N_y^2}~,
$$
which gives 
\be
{{\cal S}\over N_x} = {2\over 3}\ln N_y + c_0N_y +c_1
\label{s3}
\ee
This expression would disagree with (\ref{S2}), if $c_0$ were not zero.
It is zero, indeed. In fact:
\be
c_0 = \lim_{N_y\rightarrow\infty}{\partial\over\partial N_y}
\left({{\cal S}\over N_x}\right)
= {S(0)\over 2} +\sum_{c=1}^{\infty} S(c)
\ee
and by writing down $({{\cal S}\over N_xN_y})$ for a completely
translationally invariant two dimensional system, we obtain:
\bea
{{\cal S}\over N_xN_y} &=& {1\over 2}
\sum_{(m,l)\ne(0,0)} { l^2 - m^2 \over (l^2 + m^2)^{5/2} } = 0\nonumber\\
&=& {S(0)\over 2} +\sum_{m=1}^{\infty} S(m) = c_0\nonumber
\eea
Therefore $c_0$ vanishes and Eq.~(\ref{s3}) reduces to Eq.~(\ref{S2}).

After having discussed the shape anisotropy of an infinite stripe
$(N_x=\infty)$, upon increasing its ``thickness" $N_y$, now let us analyze
the dependence of ${\cal S}/N_xN_y$ on the aspect ratio $r$.
By handling the expression for
${{\cal S}\over N_xN_y}$ given in Appendix and by using the ``boundary"
condition ${{\cal S}\over N_xN_y}(N_x=\infty,N_y=1) = \zeta(3)$, we obtain:
\be
{{\cal S}\over N_xN_y} = \left( 1-{1\over r}\right)\left[
{2\over 3N_y}\ln N_y + {\zeta(3)\over N_y} F(r)\right]~,
\label{S(r)}
\ee
where $F(\infty)=1$ and $F(1)\approx 1$: the actual value of $F(1)$ is not 
really relevant, because the shape anisotropy vanishes in the limit $r=1$.

In Fig.~3 we compare the previous expression (as a function of $r$,
keeping fixed the value of $N_y=40$) with the exact numerical result,
obtained by exploiting Eq.~(\ref{S}). The behavior is well
reproduced by the analytical expression.

\section{Domain structures in a striped monolayer}
Now, let us turn to the study of domain structures in striped
monolayers. Our purpose is to check if $-$and when$-$ the appearance
of magnetic domains is energetically favoured. We will consider
a striped monolayer which is infinite in the $\hat x$ direction 
$(N_x=\infty)$ and we ask for which values of the parameters,
the creation of a domain wall along the $\hat y$ axis makes the system 
gain energy with respect to the collinear FM state.

We will consider two cases, according to the direction of the
magnetization: perpendicular to the striped monolayer, and in the plane,
along the ``hard" direction $(\hat y)$.
We consider just these two possibilities, because
a striped monolayer has a ``double" shape anisotropy: it has an
easy-plane effect (as in a film, or in an infinite monolayer), but
in the plane, it has also an easy-axis effect, along the $\hat x$
direction, which has been discussed in detail in the previous section. 
If the easy-plane effect is overcome by some anisotropies
$(K_\perp)$ oriented along $\hat z$, the film will break into domains.
So, our first purpose (I) will be to study the finite size effects
on this domain structure.

Conversely, if the easy-plane effect of $E_{dip}$ prevails, the
magnetization will be oriented along $\hat x$, or $-$in presence of
a sufficiently strong anisotropy $K_\parallel$ favouring the $\hat y$ axis$-$ 
along the smaller size of the stripe.
In analogy with (I), we will expect that the striped monolayer breaks into
in-plane domains. We will analyze (II) this possibility, by showing
that domains don't appear, except for very large values of the
dipolar coupling.

What we have to do $-$in cases (I) and (II)$-$ is to create a single
domain wall along the $\hat y$ axis, and to compare the domain wall
energy, with the dipolar energy gain. In the following, $w$ will
indicate the domain wall size.
For an evaluation of the dipolar term, in the Heisenberg model it will be
sufficient to consider Ising-like spins, with domains separated by an
empty region of size $w$~\cite{KP}.

In both cases (out-of-plane and in-plane domains) the domain wall
energy per unit length in the $\hat y$ direction, will be written
as $E_{dw}$ (for the Ising model: $E_{dw}=2J$, where $J$ is the
exchange coupling constant, and for the Heisenberg model:
$E_{dw}=2\sqrt{JK}$, where $K=K_\perp$ in case (I) and $K=K_\parallel$
in case (II)). 

I: out-of-plane domains. 
Because of the finite extension of the striped monolayer, the translational
invariance in the $\hat y$ direction is lost and the dipolar interaction
between a given spin and the spins of a neighbouring domain depends on
the $\hat y$ coordinate of the spin. However, for the evaluation of the
order of magnitude of the dipolar energy gain (per unit length), we can use
the following approximate formula\footnote{Indeed, it is possible to show 
that the dipolar energy gain is in between $\Delta E_{dip}(N_y)$ and
$2\Delta E_{dip}(N_y/2)$. Since $\Delta E_{dip}$ depends only
logarithmically on $N_y$, the correction is not important.},
where $-\Omega/[(x-x')^2 +y^2]^{3/2}$ is nothing but the dipolar
interaction between two spins located in $(x',0)$ and $(x,y)$, and
pointing along $\pm\hat z$:
\bea
\Delta E_{dip}(N_y)
&\approx& -2\Omega\int_{-\infty}^0 dx'\int_0^{N_y} dy\int_w^\infty
{dx\over [(x-x')^2+y^2]^{3/2}}\nonumber\\
&\approx& -2\Omega\left[\ln\left({N_y+\sqrt{N_y^2+w^2}\over w}\right)
-{\sqrt{N_y^2+w^2}-w\over N_y}\right]
\eea
In the limit $N_y\gg w$, $\Delta E_{dip}\approx -2\Omega\ln(2N_y/w)$.
The FM $\perp$ state will be destabilized if $|\Delta E_{dip}|>E_{dw}$, or
\be
N_y > {w\over 2}\exp\left({E_{dw}\over 2\Omega}\right)
\label{a2}
\ee

The explanation of this result is straightforward: out-of-plane domains
will appear only if the lateral dimensions of the striped monolayer
are larger than the size $L$ that domains would have in an infinite monolayer.
In fact, the right-hand side of Eq.~(\ref{a2}) is exactly the typical size of a domain in an infinite monolayer~\cite{YG,CV}.

II: in-plane domains, magnetized along $\pm\hat y$. 
In this case, $\Delta E_{dip}$ writes:
\bea
\Delta E_{dip}&\approx& -2\Omega\int_{-\infty}^0 dx'
\int_0^{N_y} dy\int_w^\infty
dx \left[{1\over [(x-x')^2+y^2]^{3/2}}-
{3y^2\over [(x-x')^2+y^2]^{5/2}}\right] \nonumber\\
&\approx& -2\Omega {\sqrt{N_y^2+w^2}-w\over N_y}~,
\label{II}
\eea
where the dipolar interaction $-\Omega\left\{
1/[(x-x')^2+y^2]^{3/2} -3y^2/[(x-x')^2+y^2]^{5/2}\right\}$
between two spins located in $(x',0)$ and $(x,y)$ now has a
contribution from the anisotropic part (the second term of
Eq.~(\ref{E_dip})).

In the limit $N_y\gg w$, Eq.~(\ref{II}) simply writes: 
$\Delta E_{dip}\approx -2\Omega$. 
So, the dipolar energy gain does not increase with $N_y$ and therefore the
condition for the appearance of in-plane domains ($|\Delta E_{dip}|>E_{dw}$)
can be fulfilled only if $\Omega\approx E_{dw}$, i.e. a rather large value.

\section{Conclusions}

Bulk systems differ substantially from two dimensional ones, because 
shape effects are relevant in three dimensions, but in-plane shape effects 
are negligible for stripes.  More precisely, if the magnetization lies in
the plane, its direction is mainly determined by existing magnetocrystalline
anisotropies, because dipolar shape effects rapidly vanish when the sizes
of the stripe increase, even if the aspect ratio goes to infinity.
In particular, we have shown that the shape anisotropy per spin vanishes
as ${{\ln N_y}\over {N_y}}$ upon increasing the size $N_y$, so that,
e.g. for $N_y=40$ (and $N_x=\infty$) it is reduced by a factor larger 
than 10 with respect to a single chain of spins ($N_y=1$).

The extreme weakness of shape effects in stripes has a further
consequence on the existence of in-plane domains: in fact, if
the magnetization is forced in the $\hat y$ direction (the ``hard"
direction with respect to the shape anisotropy), in-plane domains with the
magnetization alternately directed along $\pm\hat y$ appear only if 
$\Omega$ is fairly large.

We remark that our analysis has assumed that the striped
magnetic monolayer considered has a microscopic unit cell of square
symmetry, so the theory is directly applicable to epitaxial
monolayers grown on (100) substrates.
In the case of the other high symmetry orientation - the (111) one - 
the overlayer has a triangular symmetry and no in-plane anisotropy is
induced by the dipolar interaction, in the limit of an infinite
monolayer. For striped monolayers, our previous treatment should be
relevant.
Conversely, for (110) substrates the microscopic unit cell has a
rectangular symmetry, which induces a further (and possibly 
competitive) anisotropy in addition to the shape anisotropy.

\vskip 1cm
\noindent {\bf Acknowledgements -} We acknowledge Danilo Pescia for having
introduced us to the problem of the striped monolayers.

\appendix
\section{Evaluation of ${\cal S}$ in the continuum approximation}
In the continuum approximation, Eq.~(\ref{S}) for ${\cal S}$ rewrites:
\be
{\cal S} =  2I_1 -2(N_x-N_y)I_2~,
\ee
where:
\bea
I_1 &=& \int_{N_y}^{N_x-1}dx \int_{0}^{N_y-1}dy
(N_xN_y + xy -N_x y -N_y x){\cal A}(x,y)\nonumber\\
I_2 &=&  \int_{0}^{N_y-1} dx\int_{1}^{N_y-1}dy\,
y {\cal A}(x,y)\nonumber\\
{\cal A}(x,y) &=& {x^2 - y^2 \over (x^2 + y^2)^{5/2}}\nonumber
\eea

The following integrals are easily calculated~\cite{GR}:
\bea
A(x_1,x_2,y_1,y_2) &=& \int_{x_1}^{x_2}dx\int_{y_1}^{y_2}dy {\cal A}(x,y)
\nonumber\\
&=&{y_1^2-x_2^2\over 3y_1x_2\sqrt{y_1^2+x_2^2} }
+{x_2^2-y_2^2\over 3x_2y_2\sqrt{x_2^2+y_2^2} }
+ [(x_1,y_1)\leftrightarrow (x_2,y_2)]
\eea
 
\bea
B(x_1,x_2,y_1,y_2) &=& \int_{x_1}^{x_2}dx\int_{y_1}^{y_2}dy\, xy{\cal A}(x,y)
\nonumber\\
&=&{y_1^2-x_2^2\over 3\sqrt{y_1^2+x_2^2} }
+{x_2^2-y_2^2\over 3\sqrt{x_2^2+y_2^2} }
+ [(x_1,y_1)\leftrightarrow (x_2,y_2)]
\eea

\bea
C(x_1,x_2,y_1,y_2) &=& \int_{x_1}^{x_2}dx\int_{y_1}^{y_2}dy\, x{\cal A}(x,y)
\nonumber\\
&=& {1\over 3} \ln\left[
{y_1 + \sqrt{y_1^2+x_2^2}\over y_1 + \sqrt{y_1^2+x_1^2}} \right]
+ {2y_1\over 3\sqrt{y_1^2+x_2^2}} - {2y_1\over 3\sqrt{y_1^2+x_1^2}}\nonumber\\
&+& [(x_1,y_1)\leftrightarrow (x_2,y_2)]
\eea

\be
D(x_1,x_2,y_1,y_2) = \int_{x_1}^{x_2}dx\int_{y_1}^{y_2}dy\, y{\cal A}(x,y)
=-C(y_1,y_2,x_1,x_2)
\ee

In the previous expressions, 
$[(x_1,y_1)\leftrightarrow (x_2,y_2)]$ means that we have to interchange 
$x_1$ with $x_2$ and $y_1$ with $y_2$.

After some lengthy, but easy, calculations it is found that
\be
I_1 = N_xN_y A_1 + B_1 -N_x D_1 -N_y C_1~~,
\ee
where
\bea
A_1 &=& {N_x^2 -N_y^2\over 3N_xN_y\sqrt{N_x^2 +N_y^2}} - 
{\sqrt{2}\over 3N_y^2} \nonumber\\
B_1 &=& {N_x^2 -N_y^2\over 3\sqrt{N_x^2 +N_y^2}} -
{\sqrt{2}\over 3} + {N_y-N_x\over 3} \nonumber\\
C_1 &=& {1\over 3} \ln\left[ {N_x(1+\sqrt{2})\over N_y+\sqrt{N_x^2+N_y^2}}
\right] + {\sqrt{2}\over 3} - {2N_y\over 3\sqrt{N_x^2+N_y^2}}\nonumber\\
D_1 &=& {1\over 3} \ln\left[ {N_x +\sqrt{N_x^2+N_y^2)}\over
N_x(1+\sqrt{2})}\right] -  {\sqrt{2}\over 3} + 
{2N_x\over 3\sqrt{N_x^2+N_y^2}}\nonumber
\eea
and
\be
I_2 = {1\over 3}\ln\left[{1+\sqrt{2}\over 2N_y}\right] -{2-\sqrt{2}\over 3}
\ee

As a function of the previous quantities, the shape anisotropy per spin
writes:
\be
{{\cal S}\over 2N_xN_y} = {I_1\over N_xN_y} - {N_x-N_y\over N_xN_y}I_2
= A_1 +{B_1\over rN_y^2} - {D_1\over N_y} -{C_1\over rN_y} -
\left(1-{1\over r}\right){I_2\over N_y}
\ee

In the limit $N_x\rightarrow\infty$ it is immediately found that
\be
{{\cal S}\over N_xN_y} = {2\over 3N_y}\ln N_y + {\cal O}\left({1\over N_y}
\right)\label{a1}
\ee

Conversely, for any value of $r=N_x/N_y$ we obtain the following
expression:
\be
{{\cal S}\over N_xN_y} = \left( 1-{1\over r}\right)\left[
{2\over 3N_y}\ln N_y + {F'(r)\over N_y}\right]~,
\ee
where $F'(r)$ is a function which depends only very weakly on the
aspect ratio $r$: $F'(1)\approx F'(\infty)\approx 1$. 
Indeed, the term proportional to $F'(r)$ corresponds
to the term of order $(1/N_y)$ in (\ref{a1}): i.e. a term which cannot be
determined consistently by the used ``zero-order" continuum
approximation. Imposing the ``boundary condition"
$\left. {{\cal S}\over N_xN_y}\right|_{r=\infty,N_y=1} = \zeta(3)$,
corresponds to put $F'(r)=\zeta(3)F(r)$, with $F(\infty)=1$ and
$F(1)\simeq 1$.

\section{Exact calculation of ${\cal S}$ for $N_x= \infty$}

Let us consider the case of a stripe with an infinite aspect ratio
$r=N_x/N_y=\infty$, i.e. with $N_x =\infty$ and finite $N_y$. 
Since in this limit translation invariance is restored along the 
$x$ direction, it turns out that the summation in Eq.~(\ref{S1}) can be
rewritten
\begin{equation}
{{\cal S}\over N_x} = {1\over 2} \sum_{m,m'=1}^{N_y}\sum_{l=-\infty}
^{\infty} 
{ l^2 - (m-m')^2 \over \left[ l^2 + (m-m')^2 \right]^{5/2} }
= {1\over 2}N_y S(0) + \sum_{c=1}^{N_y-1} (N_y-c)S(c)
\label{B1}
\end{equation}
\noindent where $S(c)$, the interaction per unit length of two lines 
at a distance $c$, is 
\begin{equation}
S(c)=\sum_{l=-\infty}^{\infty} {{l^2 -c^2}\over 
{(l^2+c^2)^{5/2}}}
=\sum_{l=-\infty}^{\infty} {1\over {(l^2+c^2)^{3/2}}} 
-2\sum_{l=-\infty}^{\infty} {{c^2}\over {(l^2+c^2)^{5/2}}} 
\end{equation}
The self-interaction of a line, $S(0)$, is readily evaluated 
in terms of the Riemann's zeta function $\zeta(x)$
\begin{equation}
S(0)=\sum_{l=-\infty}^{\infty} {1\over {|l|^3}}=
2 \sum_{l=1}^{\infty} {1\over {l^3}}= 2 \zeta(3)=2 \cdot 1.202057
\end{equation}

To evaluate $S(c)$ for $c>0$  we use a method which was developed
by Ewald~\cite{Ewald} for converting
two-dimensional dipole sums to a rapidly converging form.
First we take into account the identity~\cite{GR}:
\begin{equation}
{1\over {\alpha^{\nu}} }={1\over { \Gamma (\nu) }} \int_0^{\infty}
dt~t^{\nu-1}~e^{-\alpha t}
\end{equation}
to rewrite $S(c)$ as follows
\begin{equation}
S(c)= {1\over {\Gamma(3/2)}} \sum_{l=-\infty}^{\infty}
\int_0^{\infty} dt~t^{1/2}~e^{-(l^2+c^2)t}
-2 c^2 {1\over {\Gamma(5/2)}}  \sum_{l=-\infty}^{\infty}
 \int_0^{\infty} dt~t^{3/2}~e^{-(l^2+c^2)t}
\end{equation}
Next we employ the identity

\begin{equation}
\sum_{l=-\infty}^{\infty} e^{-l^2 t}=\sqrt{\pi} t^{-1/2}
\sum_{n=-\infty}^{\infty} e^{-(\pi n)^2 /t}
\end{equation}

\noindent so that
\begin{equation}
S(c)=\sum_{n=-\infty}^{\infty}
\int_0^{\infty} dt~e^{-c^2 t}~ e^{-(\pi n)^2/t}~
\left\lbrack 
{{\sqrt{\pi}}\over {\Gamma(3/2)}} -2 c^2
{{\sqrt{\pi}}\over {\Gamma(5/2)}} ~ t 
\right\rbrack
\end{equation}

\noindent The integrals can be exactly evaluated in 
terms of the modified Bessel 
functions $K_{\nu}(x)$~\cite{GR}:
\begin{equation}
\int_{0}^{\infty} dt~e^{-at}~e^{-b/t} ~t^{\nu-1}=
2 \left( {b\over a} \right)^{\nu/2} K_{\nu}(2\sqrt{ab})
~~~~~~~(a,b>0)
\end{equation}

At last we obtain, for $c \ge 1$:
\begin{equation}
S(c)= {{4\pi}\over {c}} 
\sum_{n=-\infty}^{\infty}~ |n|~ K_1(2\pi c|n|) 
- {{16 \pi^2}\over 3}
\sum_{n=-\infty}^{\infty}~ n^2~ K_2(2\pi c|n|) 
\end{equation}
At this point we take into account the $ x\to 0$ expansion~\cite{Abramowitz}
$K_{\nu}(x) \simeq {1\over 2} \Gamma(\nu) ({1\over 2} x)^{-\nu}$,
so that the $n=0$ terms in the summation are found to give the
finite contributions
\bea
\lim_{n\to 0}&& n K_1(2\pi c n) = {1\over {2\pi c}}\nonumber\\
\lim_{n\to 0}&& n^2 K_2(2\pi c n) = {1\over {2\pi^2 c^2}}\nonumber
\eea
and finally we obtain for $c \ge 1$
\begin{equation}
S(c)=-{2\over {3c^2}} +{{8\pi}\over {c}} 
\sum_{n=1}^{\infty} n~K_1(2\pi c n)
-{{32\pi^2}\over 3}\sum_{n=1}^{\infty} n^2~K_2(2\pi c n)
\label{b100}
\end{equation}
The first term on the r.h.s. gives the main contribution, which
coincides with the result of the continuum limit. The two $n-$%
summations converge very rapidly because the Bessel functions 
present an exponential decay for high values of their argument:
$K_{\nu}(x) \simeq \sqrt{
{{\pi}\over {2x}}
} 
e^{-x}[ 1 +{1\over {8x}} (4 \nu^2 -1) +O({1\over {x^2}}) ]$ 
for $x \to \infty$.
In practice, excellent convergence is obtained summing about ten
terms.

\vfill\newpage

\begin{figure}
\caption{(a) Striped monolayer of size $N_x\times N_y$.
(b)~Three-dimensional slab of size $N_x\times N_y\times N_z$
(with $N_z\ll N_x,N_y$).}
\label{Fig1}
\end{figure}

\begin{figure}
\caption{Shape anisotropy per spin, ${\cal S}/(N_xN_y)$, for a striped
monolayer with infinite aspect ratio $(r=N_x/N_y = \infty)$, as a function
of $N_y$. Dots: exact numerical results, derived from 
Eqs.~(\protect\ref{B1}) and (\protect\ref{b100}).
Line: analytical approximation, Eq.~(\protect\ref{S3}).}
\label{Fig2}
\end{figure}

\begin{figure}
\caption{Shape anisotropy per spin, ${\cal S}/(N_xN_y)$, for a striped
monolayer with fixed $N_y=40$, as a function of the aspect ratio
$r=N_x/N_y$.
Dots: exact numerical results, Eq.~(\protect\ref{S}).
Dashed line: analytical approximation, Eq.~(\protect\ref{S(r)})
with $F(r)\equiv 1$.
Full line: asymptotic value (0.09153) of Eq.~(\protect\ref{S(r)}) 
for $r\to\infty$.}
\label{Fig3}
\end{figure}


\begin{thebibliography}{99}
\bibitem{Aharoni}
A. Aharoni, {\it Introduction to the theory of ferromagnetism}
(Clarendon Press, Oxford, 1996).

\bibitem{Kittel}
C. Kittel, Rev. Mod. Phys. {\bf 21}, 541 (1949).

\bibitem{HP}
T. Holstein and H. Primakoff, Phys. Rev. {\bf 58}, 1098 (1940).

\bibitem{Maleev}
S.V. Maleev, Zh. Eksp. Teor. Fiz. {\bf 70}, 2374 (1976)
[Sov. Phys. JETP {\bf 43}, 1240 (1996)].

\bibitem{YKG}
Y. Yafet, J. Kwo and E.M. Gyorgy, Phys. Rev. B {\bf 33}, 6519 (1986).

\bibitem{Allenspach}
R. Allenspach, J. Magn. Magn. Mater. {\bf 129}, 160 (1994).

\bibitem{Politi}
P. Politi. {\it Domain structures in ultrathin magnetic films}.
To appear in Comments Cond. Mat. Phys. (1997).

\bibitem{art_exp}
These structures are generally fabricated through lithographic and/or
etching techniques. See, for example:
A.O. Adeyeye, J.A.C. Bland, C. Daboo, D.G.~Hasko and H.~Ahmed, 
J. Appl. Phys. {\bf 82}, 469 (1997).

\bibitem{Abramowitz}
M. Abramowitz and I.A. Stegun (Eds.), {\it Handbook of Mathematical
Functions}, National Bureau of Standards (Applied Mathematics Series
{\bf 55}).

\bibitem{Ewald}
P.P. Ewald, Ann. Phys. {\bf 54}, 57 and 519 (1917);
{\it ibidem} {\bf 64}, 253 (1921).
See also H. Benson and D.L. Mills, Phys. Rev. {\bf 178}, 839 (1969).

\bibitem{KP}
A.B. Kashuba and V.L. Pokrovsky, Phys. Rev. Lett. {\bf 70}, 3155 (1993);
Phys. Rev. B {\bf 48}, 10335 (1993).

\bibitem{YG}
Y. Yafet and E.M. Gyorgy, Phys. Rev. B {\bf 38}, 9145 (1988).

\bibitem{CV}
R. Czech and J. Villain, J. Phys.: Condens. Matter {\bf 1}, 619 (1989).

\bibitem{GR}
I.S. Gradshteyn and I.M. Ryzhik, {\it Table of Integrals, Series,
and Products}, Academic Press (New York, 1980).
\end{thebibliography}
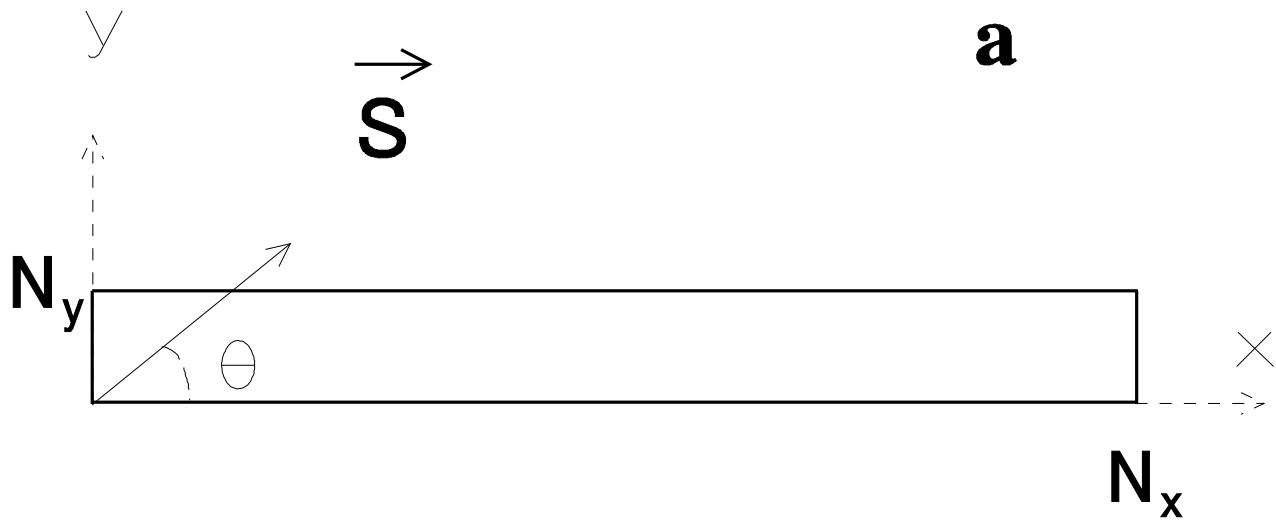
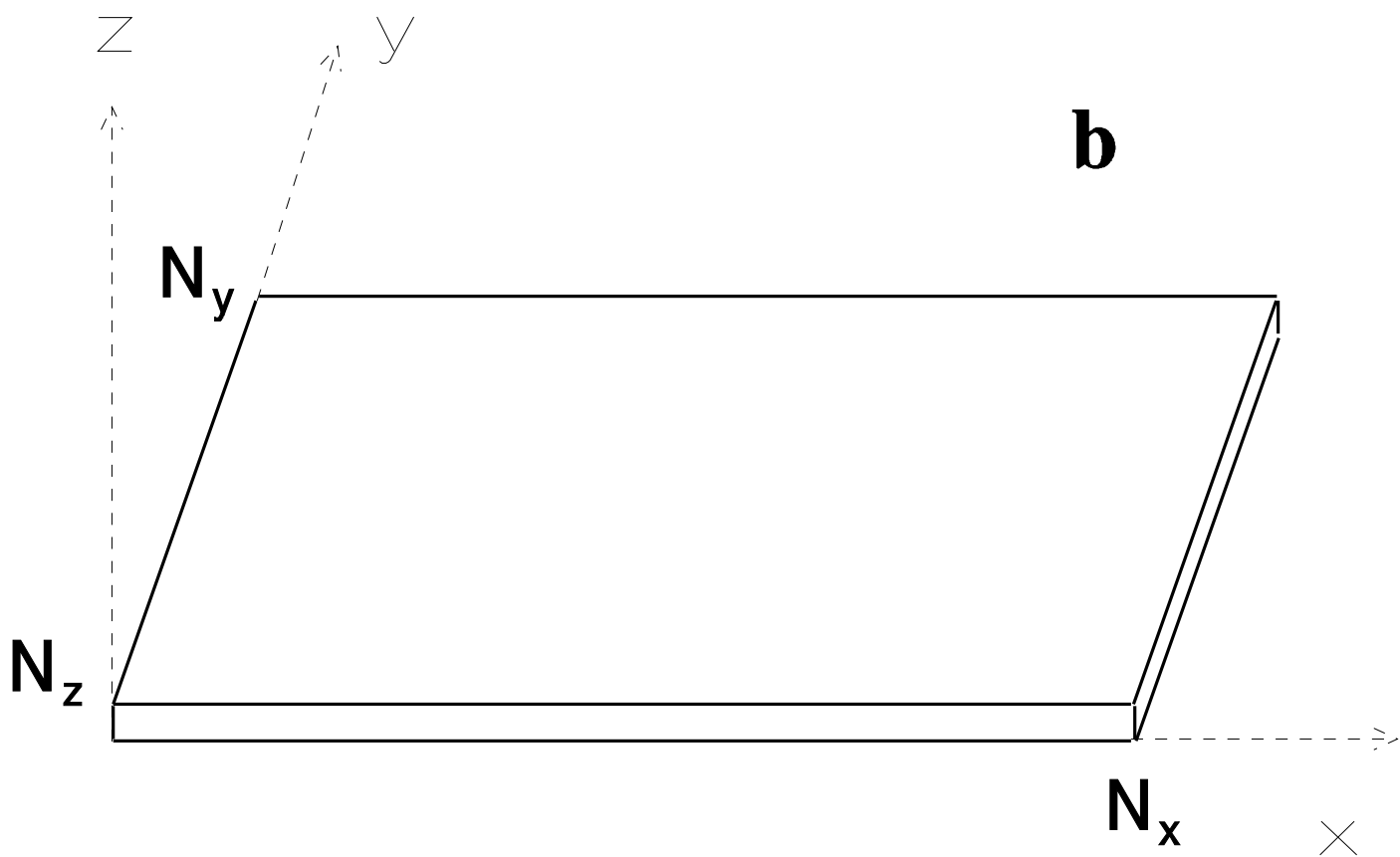
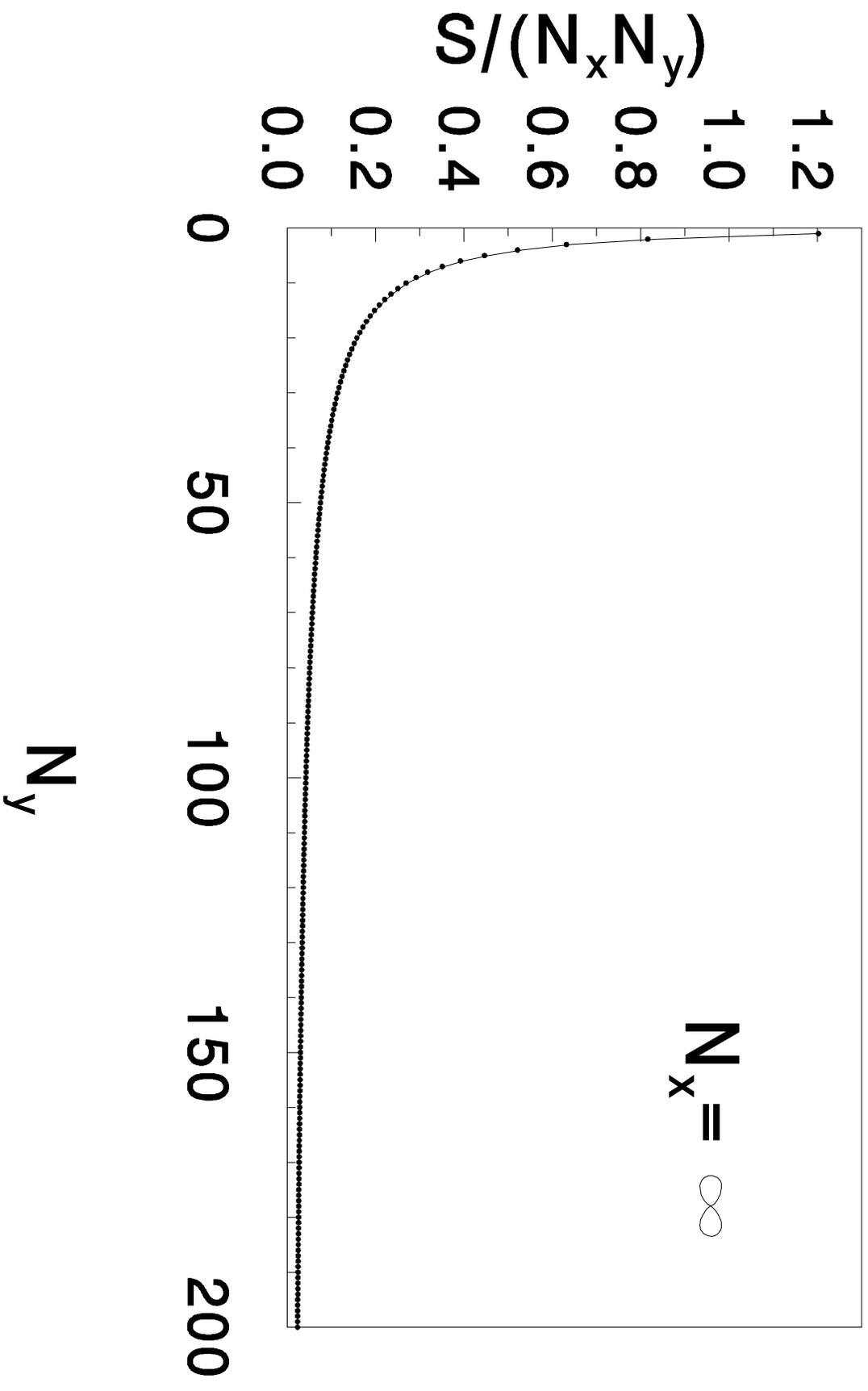
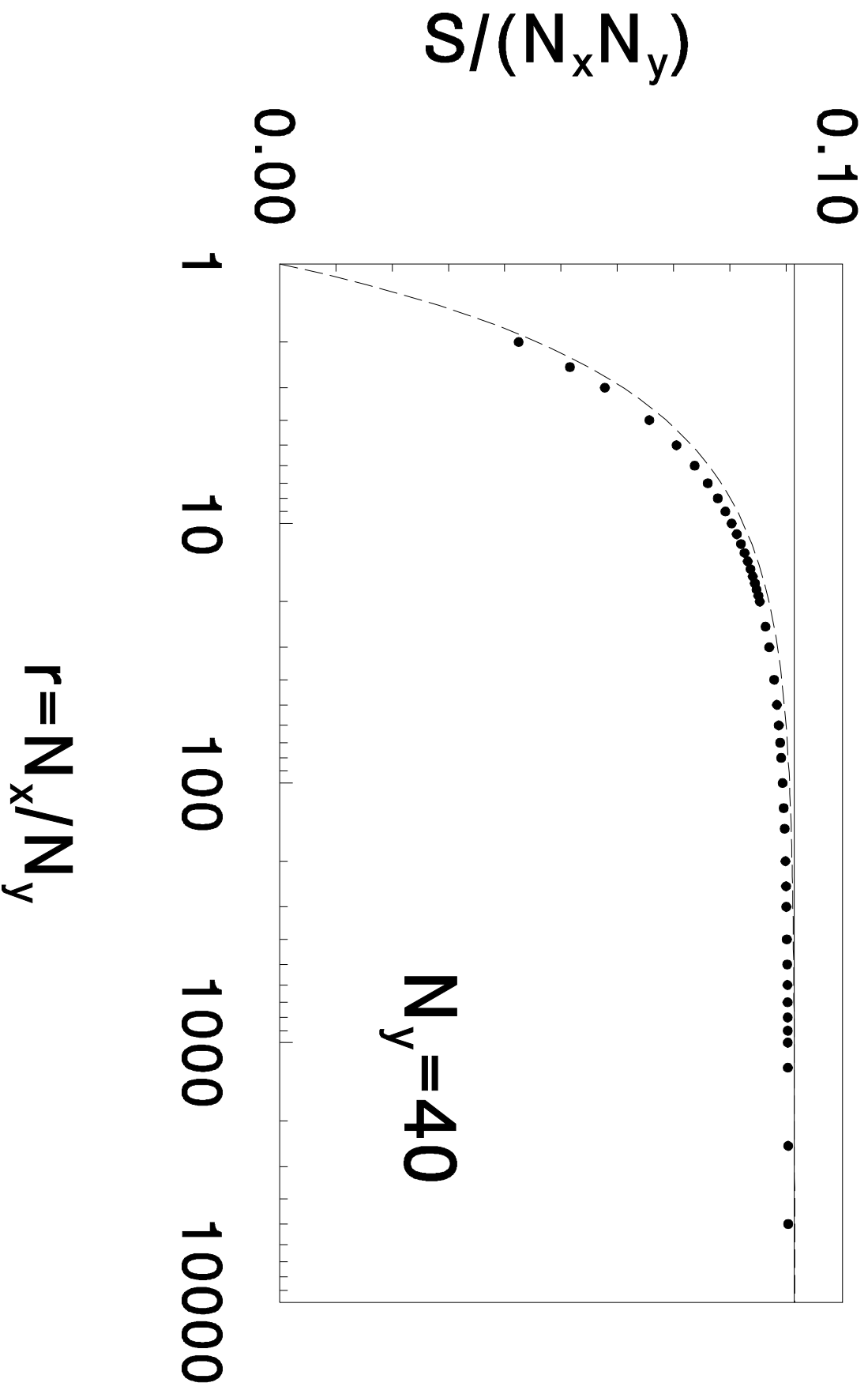
\end{document}